\documentclass{aa}  %este es la version comun a dos cols.

\usepackage{graphicx}
\usepackage{longtable}
\usepackage{txfonts}
\begin{document}

   \title{KELT-17: a chemically peculiar Am star and a hot-Jupiter planet\thanks{Based on data obtained at Complejo Astron\'omico
   El Leoncito, operated under agreement between the Consejo Nacional de Investigaciones Cient\'ificas y T\'ecnicas de la
   Rep\'ublica Argentina and the National Universities of La Plata, C\'ordoba and San Juan.}
   }

   %\subtitle{Is there a link between Am stars and planets?}
   
   \titlerunning{KELT-17: an Am star orbited by a planet}
   \authorrunning{Saffe et al.}

   \author{C. Saffe\inst{1,2,6}, P. Miquelarena\inst{1,2,6}, J. Alacoria\inst{1,6}, J. F. Gonz\'alez\inst{1,2,6},
           M. Flores\inst{1,2,6}, M. Jaque Arancibia\inst{4,5}, D. Calvo\inst{2}, E. Jofr\'e\inst{7,3,6} \and A. Collado\inst{1,2,6} 
           }

\institute{Instituto de Ciencias Astron\'omicas, de la Tierra y del Espacio (ICATE-CONICET), C.C 467, 5400, San Juan, Argentina.
               %\email{csaffe@conicet.gov.ar}
         \and Universidad Nacional de San Juan (UNSJ), Facultad de Ciencias Exactas, F\'isicas y Naturales (FCEFN), San Juan, Argentina.
         \and Observatorio Astron\'omico de C\'ordoba (OAC), Laprida 854, X5000BGR, C\'ordoba, Argentina.
               %\email{[emiliano,romina]@oac.unc.edu.ar}
         \and{Instituto de Investigaci\'on Multidisciplinar en Ciencia y Tecnolog\'ia, Universidad de La Serena, Ra\'ul Bitr\'an 1305, La Serena, Chile}
         \and Departamento de F\'isica y Astronom\'ia, Universidad de La Serena, Av. Cisternas 1200 N, La Serena, Chile.
        \and Consejo Nacional de Investigaciones Cient\'ificas y T\'ecnicas (CONICET), Argentina
        \and Instituto de Astronom\'ia, Universidad Nacional Aut\'onoma de M\'exico, Ciudad Universitaria, CDMX, C.P. 04510, M\'exico
         }

   \date{Received xxx, xxx ; accepted xxxx, xxxx}

% \abstract{}{}{}{}{} 
% 5 {} token are mandatory
 
  \abstract
   { The detection of planets orbiting chemically peculiar stars is very scarcely known in the literature. } 
   {To determine the detailed chemical composition of the remarkable planet host star KELT-17. 
   This object hosts a hot-Jupiter planet with 1.31 M$_\mathrm{Jup}$ detected by transits, being one of the more
   massive and rapidly rotating planet hosts to date.
   We aimed to derive a complete chemical pattern for this star, in order to compare it with those
   of chemically peculiar stars.}
   {We carried out a detailed abundance determination in the planet host star KELT-17 via spectral synthesis.
   Stellar parameters were estimated iteratively by fitting Balmer line profiles
   and imposing the Fe ionization balance, using the program SYNTHE together with plane-parallel ATLAS12 model atmospheres. 
   Specific opacities for an arbitrary composition and microturbulence velocity v$_\mathrm{micro}$ were calculated
   through the Opacity Sampling (OS) method. 
   The abundances were determined iteratively by fitting synthetic spectra
   to metallic lines of 16 different chemical species using the program SYNTHE.
   The complete chemical pattern of KELT-17 was compared to the recently published average pattern of Am stars.
   We estimated the stellar radius by two methods: a) comparing the synthetic spectral energy distribution with the
   available photometric data and the Gaia parallax, and b) using a Bayesian estimation of stellar parameters using stellar isochrones.
   }
   { 
   We found overabundances of Ti, Cr, Mn, Fe, Ni, Zn, Sr, Y, Zr, and Ba, together with subsolar values of Ca and Sc.
   Notably, the chemical pattern agrees with those recently published of Am stars, being then KELT-17 the first
   exoplanet host whose complete chemical pattern is unambiguously identified with this class.
   The stellar radius derived by two different methods agrees to each other and with those previously obtained in the literature.
   
   }
   {}
   
   \keywords{Stars: abundances -- 
             Stars: planetary systems -- 
             Stars: chemically peculiar -- 
             Stars: individual: {KELT-17}
            }

   \maketitle
%
%________________________________________________________________

\section{Introduction}

Classical A-type stars have elemental abundances close to solar, while Am stars present overabundances 
of most heavy elements in their spectra, particularly Fe and Ni, together with underabundances of
Ca and Sc \citep[see e.g. the recent work of ][ and references therein]{catanzaro19}.
Chemically peculiar Am stars rotate slower than average A-type stars \citep[e.g. ][]{abt00,niemczura15}, and 
most of them belong to binary systems \citep[e.g. ][]{north98,cp07,smalley14}.
These stars may host weak or ultra-weak magnetic fields driven by surface convection \citep[e.g. ][]{folsom13,blazere16,blazere20}.
The origin of their peculiar abundances is commonly attributed to diffusion processes due to gravitational settling and radiative levitation
\citep{michaud70,michaud76,michaud83,vauclair78,alecian96,richer00,fossati07},
where the stable atmospheres of slowly-rotating A-type stars would allow the diffusion processes to operate.
%On the other hand, \citet{bohm-vitense06} proposed that the accretion of interstellar material could explain the main
%characteristics of the Am stars.
%In this case, the lower convective zones of early-type stars compared to late-type stars, would favor the survival
%of accreted material in the photosphere of the star.
%The model of diffusion was also questioned by the detection of pulsations in a number of Am stars \citep{smalley11,balona11},
%although more recently \citet{smalley17} do not rule out the presence of diffusion.
%Similarly, \citet{balona15} found a rotational variability in a number of Am stars challenging the diffusion scenario,
%suggesting the accretion of metal-rich material or planets as an alternative explanation.
%Notably, \citet{stephan18} found that 70\% of planets orbiting A-type binaries\footnote{About 25\% during main-sequence and 45\% during post-main-sequence evolution}
%will be destroyed or engulfed due to the perturbation of its far stellar companion, although without a explicit reference to Am stars. 
%Then, although diffusion is not ruled out in the photospheres of Am stars, it is worthwhile to explore the possible link suggested
%between Am stars and circumstellar material or planets.

%In this context, the detection of rocky material or planets orbiting around Am stars could be a valuable piece of the puzzle. 
%However, the presence of such objects is very scarcely known. 
The detection of planets orbiting early-type stars
in general is more difficult than in late-type stars, due e.g. to the rotational broadening and lower number of spectral lines.
Just in the last few years, the detection of planets orbiting around A-type stars is slowly growing thanks mainly to detections using
transits \citep[e.g. ][]{zhou16} and direct imaging \citep[e.g. ][]{nielsen19}.
Recently, \citet{wagner16} claimed the detection of a planet by direct imaging orbiting around HD 131399 A,
an object classified as a possible Am star \citep{przybilla17}. However, this planet detection was then
ruled out and attributed to a background star \citep{nielsen17}.
Subsequent works detected planets around the likely Am star KELT-19 \citep{siverd17} and
the mild-Am star WASP-178 \citep{hellier19}.
Then, the detection of planets orbiting chemically peculiar stars is very scarcely known.

Recently, \citet{zhou16} announced the detection of a transiting hot-Jupiter planet orbiting around the early-type star \object{KELT-17}. 
The authors reported a planet with a mass, radius, and period of 1.31 M$_\mathrm{Jup}$, 1.52 R$_\mathrm{Jup}$, and 3.08 days.
This host star was only the fourth A-star with a confirmed transiting planet,
and it is one of the most massive and rapidly rotating ($\sim$44.2 km s$^{-1}$) planet hosts.
The authors adopted a metallicity of -0.018$^{+0.074}_{-0.072}$ for KELT-17 (see their Table 7), i.e. a solar or slightly subsolar metallicity for this notable star.
%However, they caution in the text (Sect. 3.2) that a possibly higher [Fe/H] could be derived by a global analysis of the transit light curves and stellar isochrones.

Currently, we have an ongoing program aimed to study [Fe/H] in early-type stars with and without planets (Saffe et al, in prep.).
The star KELT-17 is included in our sample, and a preliminary inspection of the spectra gave us the suspicion of 
peculiar abundances. This was surprising, given the previously adopted solar or subsolar [Fe/H] for this star.
Then, the exciting possibility to find a planetary host with an abnormal composition motivated us to perform a detailed chemical
analysis on this remarkable star, including different chemical species to compare with a more complete chemical pattern.
This object would be the second Am star with planets detected to date, and the first one whose
individual abundances are derived and compared to an Am pattern in detail.

%In addition, the so-called $\lambda$ B\"ootis stars are early-type objects showing underabundances ($\sim$1-2 dex) of iron-peak elements
%and near-solar abundances of C, N, O and S \citep[e.g. ][]{gray17}. Different models such as a selective accretion of
%gas separated from dust were proposed to explain this pattern \citep[see e.g. ][ and references therein]{murphy-paunzen17}.
%However, the origin of this pattern remains unclear.
%Interestingly, \citet{jura15} proposed that winds from a hot-Jupiter planet might induce this chemical pattern on its host star.
%This scenario was tested in very few objects, due to the scarcity of (early-type) hot-Jupiter planet host stars.
%This makes KELT-17 an ideal target to study through a detailed abundance analysis, which also motivates the present work.

\section{Observations}

The spectroscopic observations of KELT-17 were acquired at Complejo Astr\'onomico El Leoncito (CASLEO) between April 3 and 4, 2019.
We used the \emph{Jorge Sahade} 2.15 m telescope equipped with a REOSC echelle spectrograph\footnote{On loan from the Institute 
d'Astrophysique de Liege, Belgium}, selecting as cross disperser a grating with 400 lines mm$^{-1}$.
Three spectra of the star were obtained, followed by a ThAr lamp in order to derive an appropriate wavelength versus pixel solution.
The data were reduced using IRAF\footnote {IRAF is distributed by the National Optical Astronomical Observatories 
which is operated by the Association of Universities for Research in Astronomy, Inc., under a cooperative agreement
with the National Science Foundation.} standard procedures for echelle spectra.
The final spectra covered a visual range $\lambda\lambda$3700-6000, the resolving power R was $\sim$13000, and
the S/N per pixel measured at $\sim$5000 \AA~resulted in $\sim$350.

Achieving a proper spectrum normalization is crucial when fitting broad spectral features like \ion{H}{i} or strong \ion{Ca}{ii} lines in A-type stars.  
In echelle spectra, the normalization of orders is usually a difficult task. To obtain a reliable normalization we proceed as follows. First, we fitted
the continuum of the echelle orders without broad lines using cubic splines (6-9 pieces per order). 
Then, in problematic echelle orders, we toggled the wavelength scale to pixels and divided the observed spectrum by the continuum of an adjacent order
(or the average of both adjacent orders if available). Finally, we normalized the resulting spectrum by a very low order polynomial (order 1-3) to compensate
for small count level differences between orders.

\section{Stellar parameters and abundance analysis}

\begin{figure}
\centering
\includegraphics[width=6.5cm]{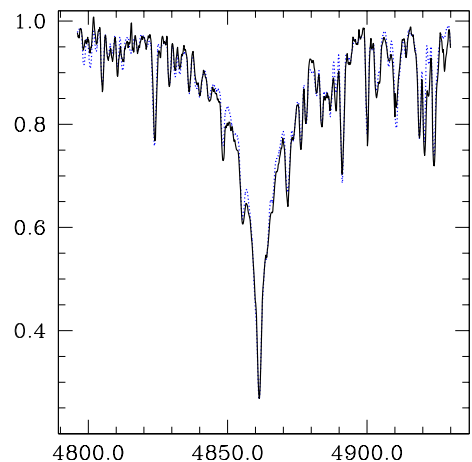}
\caption{Comparison of synthetic (blue dotted line) and observed (black continuous) spectra in
the region of the H$\beta$ line for the star KELT-17.}
\label{fig.Hbeta}%
\end{figure}

The stellar parameters T$_\mathrm{eff}$ and $\log g$ were estimated iteratively.
We fitted the observed H$\beta$ and H$\gamma$ line profiles with synthetic spectra calculated with SYNTHE \citep{kurucz-avrett81},
using ATLAS12 \citep{kurucz93} model atmospheres. Starting opacities were computed for the solar values from \citet{asplund09}.
We varied temperature and gravity adopting steps of 100 K and 0.5 dex for T$_\mathrm{eff}$ and {$\log g$}, and then we refined the grid with
steps of 1 K in T$_\mathrm{eff}$ and 0.01 dex in {$\log g$}. Synthetic spectra were convolved with a rotational profile
(using the Kurucz's command \textit{rotate}) and with an instrumental profile (command \textit{broaden}).
Balmer lines of stars cooler than $\sim$7500 K are less sensitive to gravity \citep[e.g. ][]{gray05}, then $\log g$ was also adjusted to
satisfy the ionization equilibrium of \ion{Fe}{I} and \ion{Fe}{II}.
Once determined the stellar abundances, the profiles were recomputed with specific opacities for the corresponding abundances obtained.
The final derived values are T$_\mathrm{eff}$ $=$ 7471$\pm$210 K and $\log g =$ 4.20$\pm$0.14 dex (see Table \ref{tab.params}). 
The dispersions were adopted from differences between the adjusted values for H$\beta$ and H$\gamma$, and the uncertainty of the ionization equilibrium fit.
These values are in good agreement with those obtained by \citet{zhou16}, 
T$_\mathrm{eff}$ $=$ 7454$\pm$49 K and $\log g =$ 4.220$\pm$0.023 dex.
In Fig. \ref{fig.Hbeta} we show a comparison of synthetic (blue dotted line) and observed spectra (black line) in the region of the H$\beta$ line.

\begin{table}
\centering
\caption{Stellar parameters derived in this work for the star KELT-17.}
%\hskip -0.10in
%\scriptsize
\begin{tabular}{cccc}
\hline
\hline
T$_\mathrm{eff}$ & $\log g$ & v$_\mathrm{micro}$ & $v\sin i$ \\
(K) & (dex) & (km s$^{-1}$) & (km s$^{-1}$) \\
\hline
7471$\pm$210 & 4.20$\pm$0.14 & 2.50$\pm$0.50 & 43.0$\pm$2.4 \\
\hline
\end{tabular}
\label{tab.params}
\end{table}

\begin{figure}
\centering
\includegraphics[width=8cm]{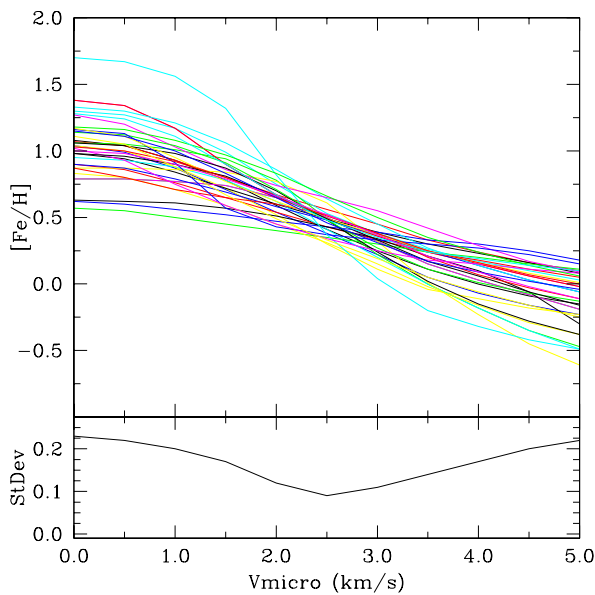}
\caption{Abundances of different \ion{Fe}{I} lines as a function of v$_\mathrm{micro}$,
showing different colors for each line. The lower panel shows the standard deviation of the different lines.}
\label{fig.vmic}%
\end{figure}

Projected rotational velocity $v\sin i$ was first estimated by fitting the observed line \ion{Mg}{II} 4481.23,
and then refined using most metallic lines in the spectra. We adopted a final value of 43.0$\pm$2.4 km s$^{-1}$,
in good agreement with the value 44.2$^{+1.5}_{-1.3}$ km s$^{-1}$ derived by \citet{zhou16}.
Then, we derived abundances using spectral synthesis.
There is evidence that Am stars possibly have a microturbulence velocity v$_\mathrm{micro}$ higher
than chemically normal A-type stars \citep[e.g. ][]{landstreet98,landstreet09}. 
Then, v$_\mathrm{micro}$ was derived by minimizing the standard deviation in the abundance of Fe lines as a function of
v$_\mathrm{micro}$.
We present in Fig. \ref{fig.vmic} the abundances of different \ion{Fe}{I} lines as a function of v$_\mathrm{micro}$,
showing different colors for each line.
In order to create this plot, v$_\mathrm{micro}$ was varied in steps of 0.5 km s$^{-1}$.
The lower panel shows the standard deviation of the different lines.
This procedure was repeated while deriving T$_\mathrm{eff}$ and {$\log g$}, estimating finally v$_\mathrm{micro} =$ 2.50$\pm$0.50 km s$^{-1}$
for the star KELT-17.

%Microturbulence velocity v$_\mathrm{micro}=$2.97$\pm$0.74 km s$^{-1}$ was estimated using the formula of \citet{gebran14},
%adopting 25\% of dispersion as they suggest.

The atomic line list and laboratory data used in this work is basically the one described in \cite{ch04},
updated with specialized references as described in the Sect. 7 of \citet{gonzalez14}.
Abundances were determined iteratively by fitting different metallic lines using the program SYNTHE.
Table \ref{tab.abunds} shows for each element the average and the total error, the number of lines, the error of the
average $\xi$\footnote{For species with one line, we adopted the average error of other elements.},
and the error when varying T$_\mathrm{eff}$, {$\log g$}, and v$_\mathrm{micro}$ by their corresponding uncertainties e$_{\Delta}$.
With the new abundance values, we derived both new opacities and a model atmosphere and restarted the process again.
In this way, abundances are consistently derived using specific opacities calculated for an arbitrary composition
using the opacity sampling (OS) method, similar to \citet{saffe18}.
Observed and synthetic spectra were compared using a $\chi^{2}$ function, the quadratic sum
of the differences between both spectra. Synthetic abundances were varied in steps of 0.01 dex until 
a minimum in $\chi^{2}$ was reached, similar to \citet{saffe-levato14}. Abundance dispersions were estimated
by quadratically adding the error of the average $\xi$ and the abundance difference when varying
T$_\mathrm{eff}$, $\log g$, and v$_\mathrm{micro}$ by their respective errors e$_{\Delta}$.

%similar to previous works \citep{saffe-levato14}.

\begin{table}
\centering
\caption{Abundances derived in this work for the star KELT-17.
The columns show the average abundance and the total error, the number of lines, the error of the
average $\xi$ and the error when varying T$_\mathrm{eff}$, {$\log g$} and v$_\mathrm{micro}$ by their corresponding uncertainties e$_{\Delta}$.}
%\hskip -0.10in
%\scriptsize
\begin{tabular}{lrrrr}
\hline
\hline
Specie     & Abundance & Number & $\xi$ & e$_{\Delta}$ \\
           & [X/H]     & of lines     &       & \\
\hline
\ion{Mg}{I}   & -0.27$\pm$0.17 & 3  & 0.10 & 0.13 \\
\ion{Mg}{II}  & -0.10$\pm$0.35 & 1  & 0.11 & 0.33 \\
\ion{Si}{II}  &  0.05$\pm$0.44 & 2  & 0.43 & 0.11 \\
\ion{Ca}{II}  & -0.99$\pm$0.16 & 1  & 0.11 & 0.11 \\
\ion{Sc}{II}  & -0.88$\pm$0.19 & 2  & 0.07 & 0.17 \\ 
\ion{Ti}{II}  &  0.21$\pm$0.22 & 18 & 0.05 & 0.21 \\
\ion{Cr}{II}  &  0.58$\pm$0.13 & 17 & 0.05 & 0.12 \\
\ion{Mn}{I}   &  0.19$\pm$0.15 & 6  & 0.09 & 0.12 \\
\ion{Fe}{I}   &  0.46$\pm$0.20 & 35 & 0.02 & 0.20 \\ 
\ion{Fe}{II}  &  0.49$\pm$0.15 & 25 & 0.03 & 0.15 \\
\ion{Ni}{II}  &  0.56$\pm$0.18 & 1  & 0.11 & 0.14 \\
\ion{Zn}{I}   &  0.39$\pm$0.17 & 2  & 0.01 & 0.16 \\ 
\ion{Sr}{II}  &  0.94$\pm$0.27 & 1  & 0.11 & 0.24 \\
\ion{Y}{II}   &  1.29$\pm$0.42 & 5  & 0.11 & 0.41 \\
\ion{Zr}{II}  &  0.12$\pm$0.21 & 1  & 0.11 & 0.18 \\ 
\ion{Ba}{II}  &  1.47$\pm$0.36 & 2  & 0.25 & 0.26 \\ 
\hline
\end{tabular}
\label{tab.abunds}
\end{table}

\section{Discussion}

\begin{figure}
\centering
\includegraphics[width=6.5cm]{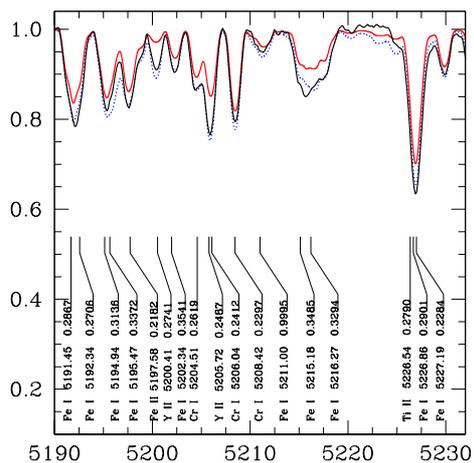}
\caption{Observed (black line) and synthetic (red and blue) spectra for the star KELT-17 between 5190 and 5230 \AA.
Red and blue colors correspond to solar and the derived abundances.}
\label{fig.region}%
\end{figure}

We present in Fig. \ref{fig.region} an example of observed and synthetic spectra
of the star KELT-17, for a region between 5190 and 5230 \AA. The observed spectrum is shown in black,
while synthetic spectra are plotted in red (solar composition) and blue dotted lines (adopted composition).
We include the identification of the lines together with their intensities (between 0 and 1).
It is clear from this plot that KELT-17 does not match solar abundances: most metallic lines
such as \ion{Fe}{II} and \ion{Cr}{II} are more intense in KELT-17. This is also evident in lines such
as \ion{Y}{II} 5200.41 \AA, which is clearly present in KELT-17 and very weak adopting solar values.

\begin{figure}
\centering
\includegraphics[width=6.5cm]{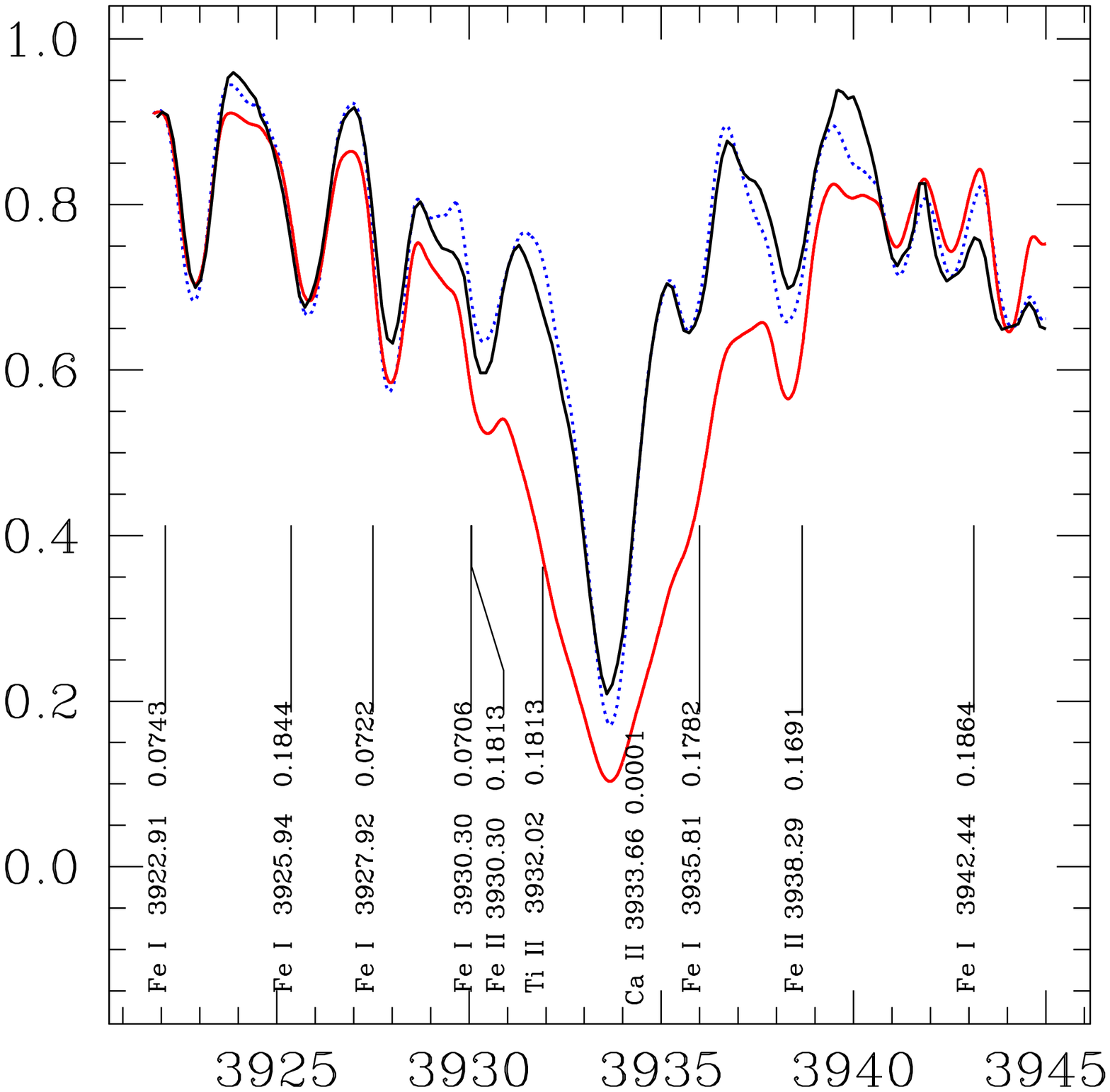}
\caption{Observed and synthetic spectra of KELT-17 in a region near the line \ion{Ca}{II} 3933.68 \AA.
Line colors used are similar to the Fig. \ref{fig.region}.}
\label{fig.CaII}%
\end{figure}

\begin{figure}
\centering
\includegraphics[width=6.5cm]{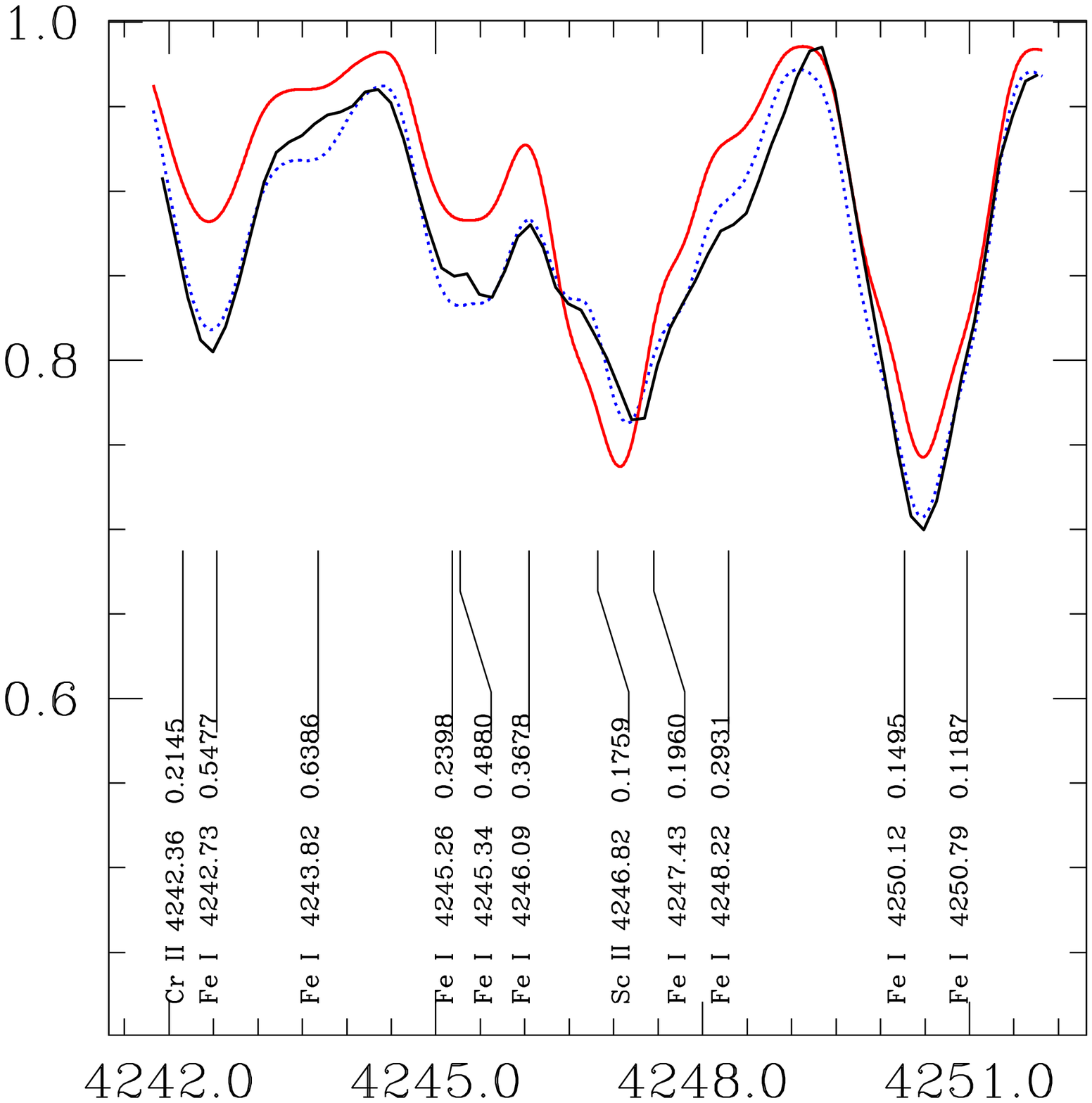}
\caption{Observed and synthetic spectra of KELT-17 in a region near the line \ion{Sc}{II} 4246.82 \AA.
Line colors used are similar to the Fig. \ref{fig.region}.}
\label{fig.ScII}%
\end{figure}

\begin{figure}
\centering
\includegraphics[width=6.5cm]{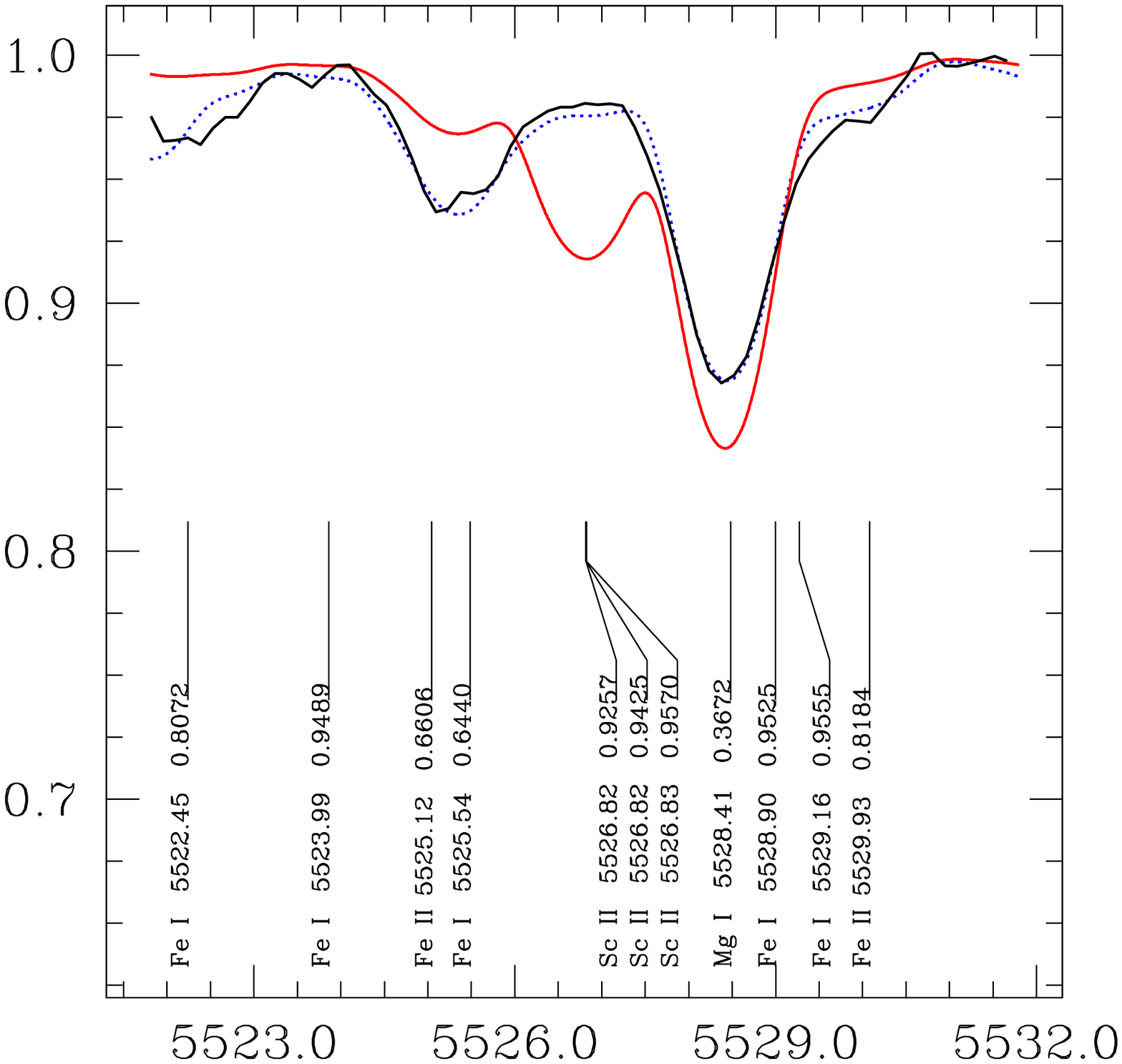}
\caption{Observed and synthetic spectra of KELT-17 in a region near the line \ion{Sc}{II} 5526.82 \AA.
Line colors used are similar to the Fig. \ref{fig.region}.}
\label{fig.ScII.2}%
\end{figure}

In addition to overabundances, the chemical pattern of Am stars is also characterized by subsolar values
of Ca and Sc \citep[see e.g. ][]{catanzaro19}. 
We present in Figs. \ref{fig.CaII} and \ref{fig.ScII} observed and synthetic spectra
of KELT-17 near the lines \ion{Ca}{II} 3933.68 \AA~and \ion{Sc}{II} 4246.82 \AA, using the same notation.
Notably, the \ion{Ca}{II} line in KELT-17 (black line) is much weaker than in a solar composition
spectrum (red line).
Similarly, in Fig. \ref{fig.ScII} we can see that the lines of \ion{Fe}{II}
and \ion{Cr}{II} are more intense in KELT-17, while the line \ion{Sc}{II} 4246.82 \AA~is weaker than
a supposed solar-like composition. A more clear example of the weakness of \ion{Sc}{II}
in the star KELT-17 can be seen in Fig. \ref{fig.ScII.2}. In this case,
we plotted a region near the line \ion{Sc}{II} 5526.82 \AA. A solar-like composition should
clearly present this line (red line), while in the spectrum of KELT-17
this line is almost absent (black line).

\begin{figure}
\centering
\includegraphics[width=7.5cm]{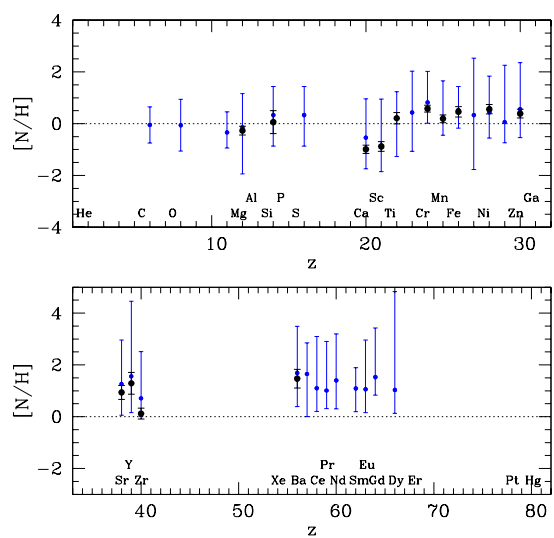}
\caption{Chemical pattern of KELT-17 (black) and average pattern of Am stars (blue).
Upper and lower panels correspond to z$<$32 and z$>$32.}
\label{fig.pattern}%
\end{figure}

Overabundances and subsolar values of species showed in Figs. \ref{fig.region} to \ref{fig.ScII.2}
are similar to Am stars.
A more quantitative comparison can be performed by directly comparing the 
abundance pattern of KELT-17 with those of Am stars in general.
In Fig. \ref{fig.pattern} we show the chemical pattern
as abundance versus atomic number for the different chemical species.
For the abundance pattern of Am stars (blue color in Fig. \ref{fig.pattern}),
we used the average values recently derived by \citet{catanzaro19} from a sample of 62 Am stars,
where vertical bars correspond to their maximum and minimum values.
The abundance values derived for the star KELT-17 are shown in black.
These values correspond to 16 different species identified in the spectrum: 
\ion{Mg}{I}, \ion{Mg}{II}, \ion{Si}{II}, \ion{Ca}{II}, \ion{Sc}{II}, \ion{Ti}{II}, \ion{Cr}{II},
\ion{Mn}{I}, \ion{Fe}{I}, \ion{Fe}{II}, \ion{Ni}{II}, \ion{Zn}{I}, \ion{Sr}{II}, \ion{Y}{II},
\ion{Zr}{II} and \ion{Ba}{II}.
From this plot, we see that KELT-17 agrees in general with the chemical characteristics
of Am stars.

\begin{figure}
\centering
\includegraphics[width=8.0cm]{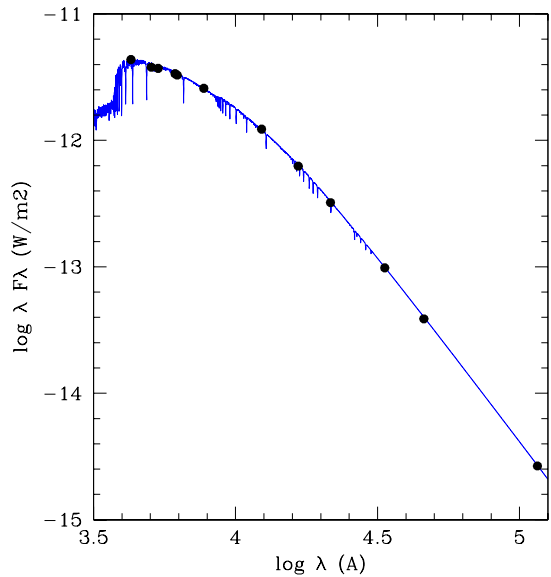}
\caption{Spectral energy distribution computed with ATLAS12 model atmospheres (blue line) and
available photometry in different bands (black circles).}
\label{fig.sed}%
\end{figure}

We present in Fig. \ref{fig.sed} the spectral energy distribution computed with ATLAS12 model atmospheres and
the available photometry in different bands, adopting a Gaia parallax of 4.387$\pm$0.048 mas \citep[][]{gaia}.
We used the B$_{T}$ and V$_{T}$ TYCHO-2 magnitudes \citep{hog00}, 
G, G$_{BP}$ and G$_{RP}$ Gaia DR2 magnitudes \citep{gaia},
The Amateur Sky Survey V band \citep[TASS, ][]{droege06},
the Carlsberg Meridian Telescope survey r' band \citep{evans02},
2MASS J, H and Ks bands \citep{cutri03}, and WISE magnitudes \citep{cutri12}.
The reddening was calculated according to the distance and position using the extinction maps of
\citet{schlegel98}\footnote{https://irsa.ipac.caltech.edu/applications/DUST/}, following the procedure of \citet{bilir08}.
From this plot, we estimated a stellar radius of 1.697$\pm$0.063 R${_\sun}$.
We also derived the radius through a Bayesian analysis using the PARSEC stellar isochrones \citep[][]{bressan12},
from the T$_\mathrm{eff}$, an apparent magnitude of V$=$9.23$\pm$0.02 from the TYCHO-2 catalog \citep{hog00},
and the Gaia parallax. In this way, we obtained R $=$ 1.680$\pm$0.092 R${_\sun}$ for the star KELT-17.
We adopted a solar global composition when using isochrones with this chemically peculiar star,
similar to previous works \citep[e.g. ][]{pohnl04,kochukhov-bagnulo06}.
Both estimations of the stellar radius agree with the previous determination of \citet{zhou16},
who derived R $=$ 1.645$^{+0.060}_{-0.055}$ R${_\sun}$. 
Then, the planetary radius previously estimated using the transit depth does not change significantly.

%Then, adopting a planet-to-stellar ratio 
%R$_{p}$/R $\sim$0.09526$^{+0.00088}_{-0.00085}$ from \citet{zhou16}, we estimate a planet
%radius of R$_{p}$ 1.610$\sim$ R$_\mathrm{Jup}$,
%in agreement with 1.525$^{+0.065}_{-0.060}$ derived by \citet{zhou16}.}}

\section{Summary}

We performed a chemical analysis of the exoplanet host star KELT-17 and found overabundances of Ti, Cr, Mn,
Fe, Ni, Zn, Sr, Y, Zr, and Ba, together with subsolar values of Ca and Sc.
The chemical pattern agrees with those recently published of Am stars, being then KELT-17 the first
exoplanet host whose complete chemical pattern is unambiguously identified with this class.
We also derived the stellar radius by two different methods, obtaining good agreement between them
and with those previously derived in the literature.
Therefore, the classification of KELT-17 as an Am star has no significant impact on the corresponding planet parameters.

\begin{acknowledgements}
We thank the anonymous referee for suggestions that greatly improved the paper.
The authors thank Dr. R. Kurucz for making their codes available to us.
CS, MF, and JFG acknowledge financial support from the CONICET of Argentina through grant PIP 0331.
MJA acknowledges the financial support of DIDULS/ULS, through the project PI192135.
\end{acknowledgements}

\end{document}